\documentclass{fundam}

\usepackage{url} 
\usepackage[ruled,lined]{algorithm2e}
\usepackage{graphicx}
\usepackage{tikz}
\usetikzlibrary{automata, positioning, arrows}
\usepackage{svg}
\usepackage{subcaption}
\usepackage{multirow}

\begin{document}

\title{Hyperspectral Unmixing of Agricultural Images taken from UAV Using Adapted U-Net Architecture}

\address{Vytautas Paura, Institute of Data Science and Digital Technologies, Vilnius University}

\author{Vytautas Paura\\
Institute of Data Science and Digital Technologies\\
Vilnius University \\
Vilnius, Lithuania \\
vytautas.paura@mif.stud.vu.lt
\and Virginijus Marcinkevičius\\
Institute of Data Science and Digital Technologies\\
Vilnius University\\
Vilnius, Lithuania \\
virginijus.marcinkevicius@mif.vu.lt
}

\maketitle

\runninghead{Vytautas Paura, Virginijus Marcinkevičius}{Hyperspectral Unmixing of Agricultural Images}

\begin{abstract}
The hyperspectral unmixing method is an algorithm that extracts material (usually called endmember) data from hyperspectral data cube pixels along with their abundances. Due to a lower spatial resolution of hyperspectral sensors data in each of the pixels may contain mixed information from multiple endmembers. In this paper we create a hyperspectral unmixing dataset, created from blueberry field data gathered by a hyperspectral camera mounted on a UAV. We also propose a hyperspectral unmixing algorithm based on U-Net network architecture to achieve more accurate unmixing results on existing and newly created hyperspectral unmixing datasets.
\end{abstract}

\begin{keywords}
Hyperspectral Unmixing, Remote Sensing, Deep Neural Networks, Hyperspectral Dataset.
\end{keywords}

\section{Introduction}

The growing popularity of remote sensing systems and advancements in hyperspectral imaging technologies created a growing interest in using these technologies in various agricultural applications. Most commonly used near-infrared hyperspectral cameras enable the collection of large amounts of spatial and spectral information simultaneously. A large amount of spectral data comes at the cost of spatial resolution compared to RGB or multispectral cameras. Smaller spatial resolution leads to multiple material data being mixed in each pixel of hyperspectral data, especially when using satellite hyperspectral sensors. To solve this problem of mixed data, hyperspectral unmixing algorithms are used, which usually solve three tasks at the same time: finding material (called endmembers) counts, calculating material spectral signatures, and finding mixture amounts (abundances) in each hyperspectral pixel. In this paper we concentrated on expanding the hyperspectral unmixing research on agricultural hyperspectral data and compare our proposed algorithm to a transformer based algorithm. This paper is structured in few chapters: literature review about the hyperspectral unmixing algorithms and thei usage on agricultural data,

\section{Related work}
\par
This section describes works related to hyperspectral unmixing use cases with a focus on agricultural data.

\subsection{Hyperspectral unmixing algorithms}
This section describes a few of the most common types of hyperspectral unmixing algorithms. These three types of algorithms are \cite{informatica}: 
\begin{itemize}
    \item Semi-supervised sparse regression modelling.
    \item Unsupervised non-negative matrix factorization methods.
    \item Unsupervised deep learning autoencoder neural networks.
\end{itemize}

Sparse regression algorithms are used due to the fact that in a hyperspectral image most of the pixels will have only a few material data mixed inside compared to all of the endmembers in the image, which in turn creates an abundance matrix that is sparse.

Non-negative matrix factorization algorithms are used because the information gathered by the hyperspectral sensors can never be negative, and the hyperspectral cube can be factored into abundance, endmember and resodual noise matrices.

The last type of the algorithms are neural networks, specifically autoencoder type networks, that create an artificial neuron bottleneck to compress the data into a latent space, extracting spatial and spectral features from hyperspectral images. Deep neural network algorithms are now becoming the most popular due to their ability to learn from complex nonlinear data. Due to the base model architecture design, these types of neural networks can be trained using the difference between the original image and its reconstruction, making the algorithm unsupervised.

\subsection{Agricultural hyperspectral data unmixing}
A comprehensive and extensive review paper on hyperspectral data usage and unmixing in agriculture was written by Geurri et al. \cite{GUERRI2024100062}. The authors explore the use cases of hyperspectral data in agriculture and available algorithms used to unmix this hyperspectral data. They explore these types of algorithms in their paper: Autoencoder denoising \cite{10225536}, Convolutional Neural Networks \cite{8082108}, Recurrent Neural Networks for classification \cite{ZHOU201939}, Deep Belief Networks \cite{rs14061484}, Generative Adversarial Networks for denoising \cite{rs14081790} and super-resolution \cite{9837938}, Transfer Learning for classification tasks \cite{isprs-archives-XLIII-B3-2022-405-2022}, Semi-Supervised Learning classification \cite{app12083943}, Unsupervised learning classification \cite{8082108}. From their paper, a conclusion can be made that the most popular unmixing methods recently are all in the domain of deep learning algorithms.

A paper by Sangeetha Annam and Anshu Singla \cite{9702646} uses supervised and unsupervised machine learning methods to detect heavy metals (arsenic (As), cadmium (Cd), and lead (Pb)) in soil from hyperspectral data. With unsupervised k-means algorithm achieving the best resulting accuracy of around 98\%.

\section{Hyperspectral dataset of blueberry fields}
\par
\begin{figure}[t]
\includegraphics[width=\textwidth]{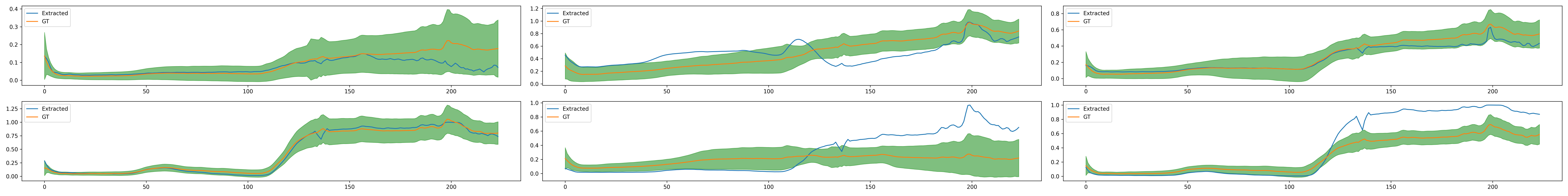}
\caption{Model predicted endmember (blue) comparison to ground truth endmember averages (orange) and their variations (green) for each of the six classes.} \label{fig:endmembers}
\end{figure}

In this section, we describe the field data gathering process and the creation of a hyperspectral unmixing dataset from hyperspectral data gathered by a UAV flying over a blueberry field. Expanding on the review paper by Geurri et al. \cite{GUERRI2024100062} and our previous work \cite{informatica}, a conclusion was made that the amount of openly available hyperspectral data, especially in agricultural areas, is limited. In this paper, we create a hyperspectral unmixing dataset from UAV data gathered in blueberry fields. Blueberry field dataset is an expansion on our previous work \cite{10.1007/978-3-031-63543-4_13} with these key differences:
\begin{itemize}
    \item Further research on classification accuracy using the VCA algorithm, changing the variation threshold in class data sampling to 1.5 $\sigma$ from 2.
    \item Expansion of the dataset from single hyperspectral cube to 3 cubes, for more extensive experimentation and data variety.
    \item Experimentation on classification, with the best results achieved by keeping the same class distribution in all of the hyperspectral data cubes.
    \item Extraction of class variation for each class for better algorithm accuracy assessment during development. The additional experiment used was to check if any of the endmembers guessed by the model were inside the class variation instead of checking only the RMSE value. Example image is given in fig \ref{fig:endmembers}. RMSE values are calculated from the endmember averages and will be higher than zero while the predicted endmember may still be within the class variance across all of the spectral bands.
\end{itemize}

\subsection{Data gathering}
\par
Raw hyperspectral data was gathered using an \textit{Aurelia} X6 drone \cite{drone} with \textit{SPECIM} hyperspectral push broom camera \cite{SPECIM} flying over a blueberry field. Push broom hyperspectral camera records the data of all (in case of this data gathering mission) 224 spectral bands in lines of 1024 pixels wide. In turn, final hyperspectral cube size gathered depends on the flight length but will always have a similar size of 1024 pixels wide with 224 spectral bands gathered. Data gathering flight was conducted from 70 meters above ground, and a required drone speed was calculated from altitude to keep the pixels square. The final pixel size of gathered UAV hyperspectral cubes is 5 x 5 centimeters. 
\par
To keep the data as accurate as possible, the data recording has to be done on straight flight paths only, called flight lines. Each flight line creates a separate hyperspectral data cube of size 1024 * x * 224, where x depends on the line length and camera recording speed. An exposure time of 6 ms and the camera fps (or lines per second in push broom camera case) set to 100. 

\subsection{Calibration}
\par
 To keep the data consistent and comparable between flights, a set of calibration carpets was deployed in the field with their laboratory-calibrated reflectance values of 5\%, 10\%, and 40\%, and a data cube with the camera lens closed to gather fully dark data or sensor noise. Calibration was performed using the reference/reflectance carpets placed in the field with one of the drone flight lines intersecting the carpets. The main methodology used was from the article by James Burger and Paul Geladi \cite{calibration}.

\subsection{Raw hyperspectral data}
\par
From the multiple hyperspectral data cubes gathered during the UAV mission, three data cubes were selected as the base of the unmixing dataset. Three cubes were selected in order to create train, test, and validation data cubes, respectively. All of the data cubes share the same set of endmembers (e.g. blueberries, grass, soil, water, areas obscured by shadows), but the data was collected over the field at different times and in different places of the same blueberry field. Three cubes are used to increase data variety and, in turn, check algorithm robustness to changes in field data. The three data cubes have these parameters:
\begin{itemize}
    \item Cube 1 shape: 1024 pixels wide, 3177 pixels long with 224 spectral bands of depth.
    \item Cube 2 shape: 1024 pixels wide, 3047 pixels long with 224 spectral bands of depth.
    \item Cube 3 shape: 1024 pixels wide, 2815 pixels long with 224 spectral bands of depth.
    \item All cubes have the same spectral data collected ranging 400 to 1000 nm with an average distance of between bands of approximately 2.5 nm.
\end{itemize}

\par
Hyperspectral cube RGB representations created by data integration over CIE 1931 XYZ color matching functions and conversion from XYZ to RGB are given below in figure \ref{fig:cubes}. 

\begin{figure}
\centering
  \begin{subfigure}[b]{\textwidth}
    \includegraphics[width=\textwidth]{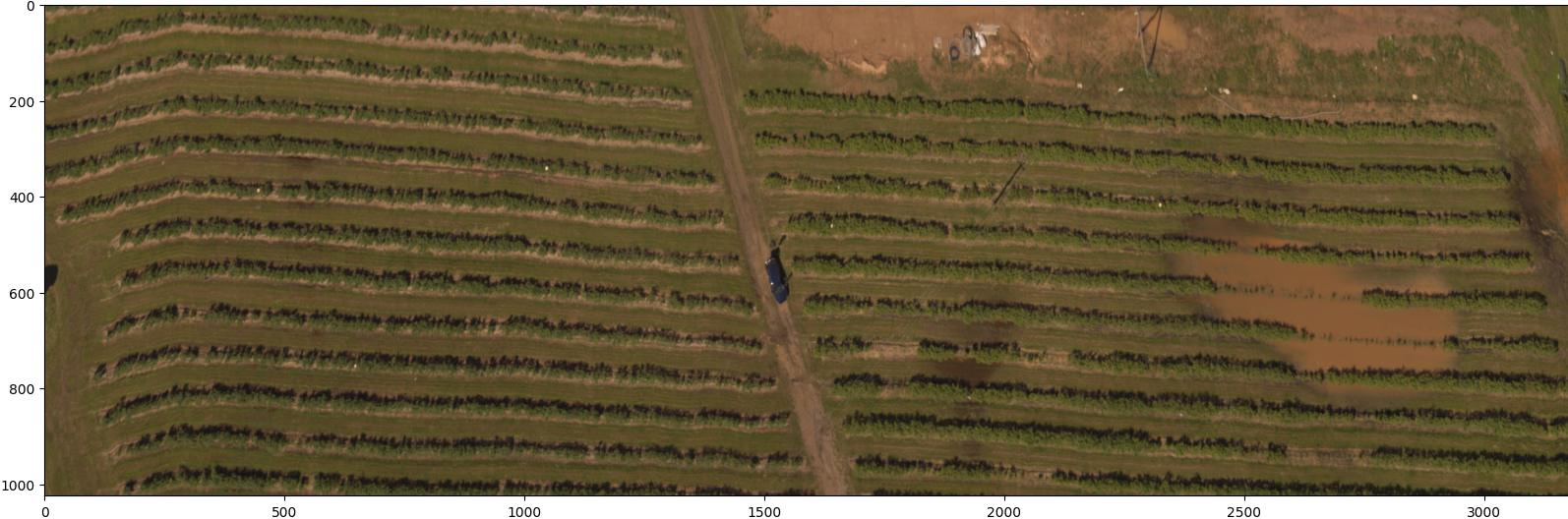}
  \end{subfigure}
 \vfill
  \begin{subfigure}[b]{\textwidth}
    \includegraphics[width=\textwidth]{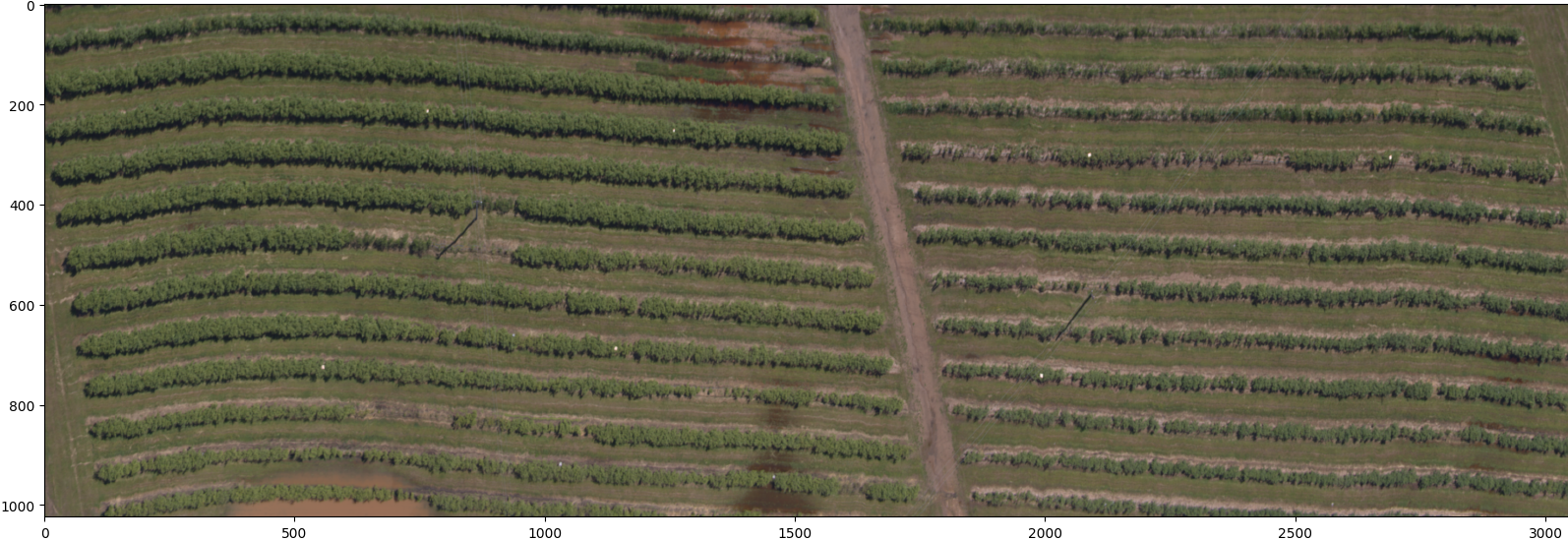}
  \end{subfigure}
\vfill
  \begin{subfigure}[b]{\textwidth}
    \includegraphics[width=\textwidth]{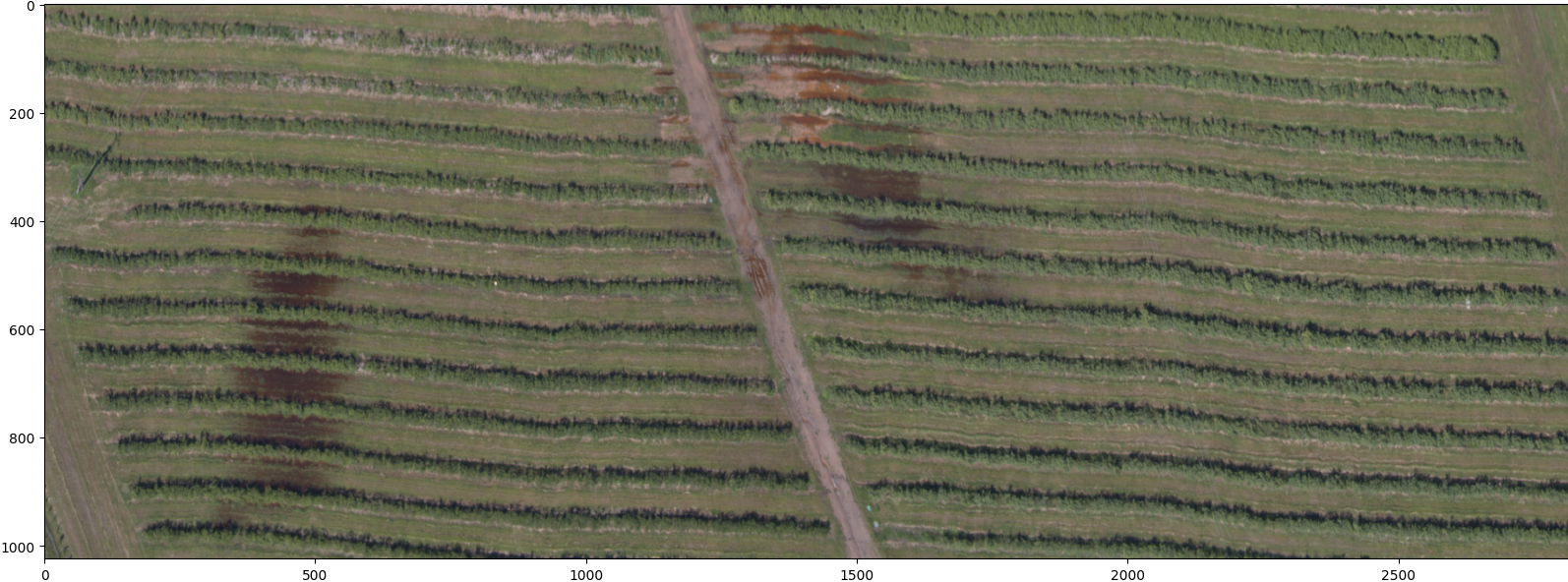}
  \end{subfigure}
\caption{RGB representation of the hyperspectral blueberry data cubes, from top to bottom cubes 1, 2 and 3 are shown.}
\label{fig:cubes}
\end{figure}

\subsection{Hyperspectral dataset ground truth creation}
\par
\begin{figure}[t]
\includegraphics[width=\textwidth]{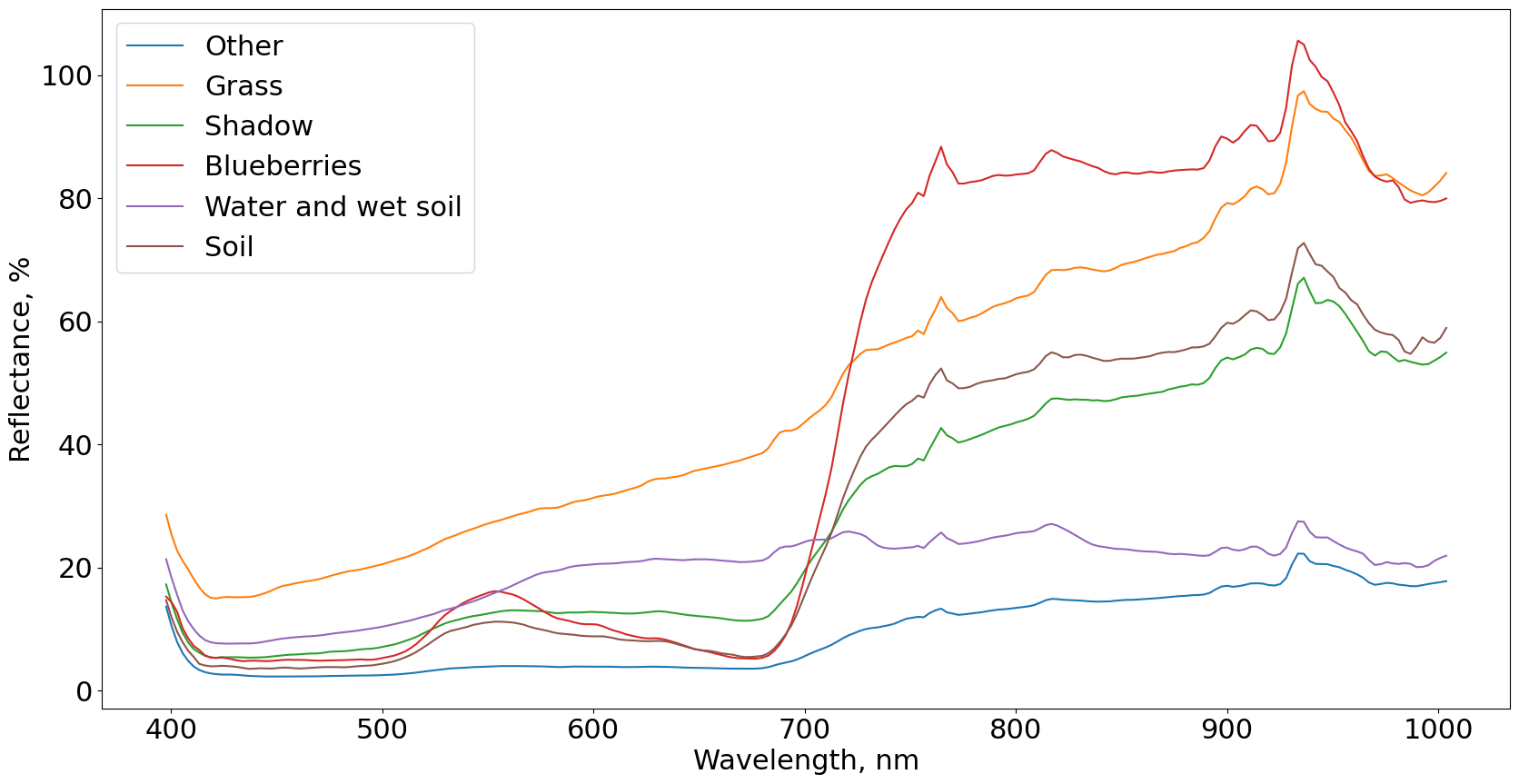}
\caption{Averages, for each of the six classes, of extracted endmembers used as the ground truth for the created hyperspectral unmixing dataset.} \label{fig:classes}
\end{figure}

Gathered field hyperspectral data has a disadvantage over laboratory data in that completely accurate classification and ground truth data are not available and are difficult to create. To apply as accurate as possible data classification a collection of classes were extracted from the data cubes using an unsupervised method called Vertex Component Analysis (VCA) \cite{vca}. The suggested endmembers extracted using the VCA algorithm were used as the ground truth classes for this hyperspectral unmixing dataset. By selecting multiple endmember counts, the algorithm extracted possible endmembers from the hyperspectral data cube (Cube 1 was used for extraction). A higher amount of classes resulted in a smaller sample size for each class and an increase in unmixing difficulty down the line. Six classes were selected by the VCA to maintain a high-class representation of the hyperspectral image and a reasonable calculation difficulty. Raw class data is represented in fig \ref{fig:classes}. Classes represent blueberry crops, bare soil, grass, data in shadow, water, and other data separate from other classes.

\par
By using the extracted possible endmembers whole data cube classification can be performed. To classify the data, each pixel was checked against the endmembers, and RMSE \ref{eq:rmse} values were calculated for each one. The class, selected for each pixel, was one with the lowest RMSE value to one of the six extracted endmembers. To keep a variation of data in the hyperspectral data cube, each pixel was left as the original data if it was within 1.5 $\sigma$ variation in its respectable class. In other words, the pixels close to the extracted endmember were left unchanged and only given a class number. Pixels out of 1.5 $\sigma$ variation were replaced by a random pixel data from within the class data distribution, to keep the computation times shorter and spatial variation higher.
\par
\begin{figure}[t]
\includegraphics[width=\textwidth]{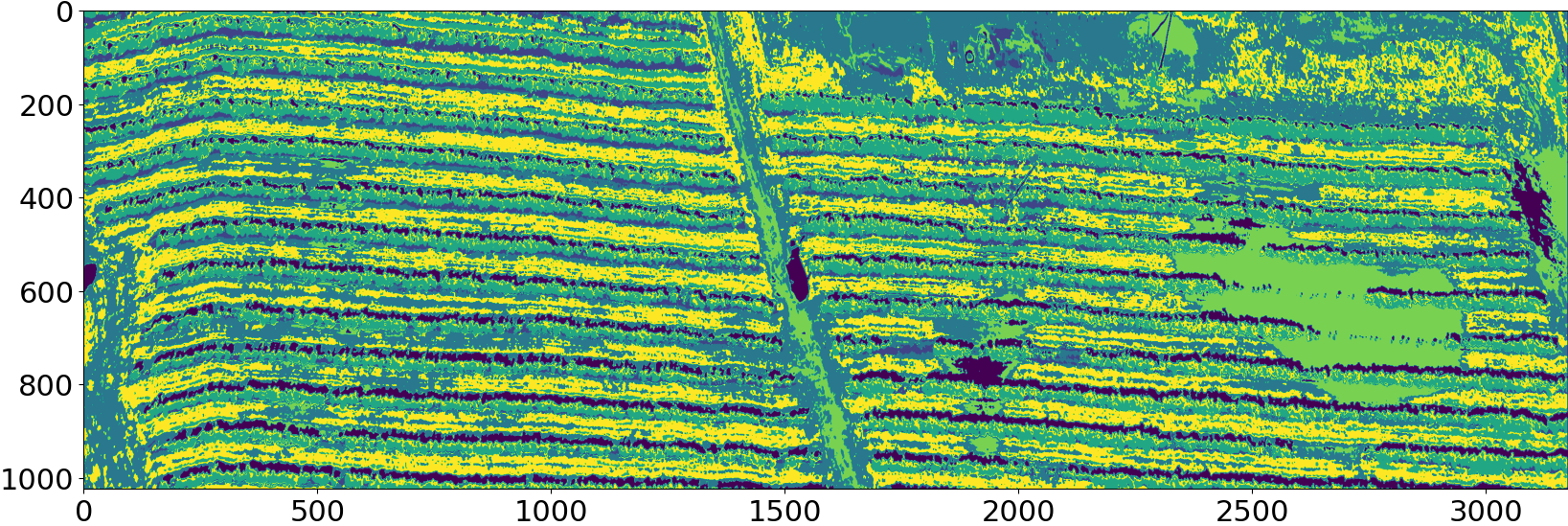}
\caption{Class distribution in the hyperspectral Cube 1. Color class representation: \color{yellow}Yellow - bare soil; \color{green}Green - Blueberries; \color{blue}Blue - Grass; \color{teal}Dark blue - Shadowed data; \color{lime}Light green - Water and wet soil; \color{black}Black - Other data} \label{fig:clss}
\end{figure}

Other endmember extraction and cube classification methods may be used, these methods were used to balance the resource requirements, labeling time and data variation inside the dataset. Full raw and classified data is published as open data for use in other experiments and hyperspectral unmixing and classification tasks.
Data is published on Zenodo platform with open access: 
\url{https://doi.org/10.5281/zenodo.13856357}

Class distribution on the hyperspectral data Cube 1 is shown in fig \ref{fig:clss}

\subsection{Ground truth data mixing for hyperspectral unmixing dataset creation}
Classified hyperspectral data cubes were mixed using a sliding window and linear mixing. Multiple sliding window sizes were considered, and experiments were conducted on window kernel sizes of 2, 3, and 4. The main kernel size used was 3, to keep a balance between the amount of data mixed (9 pixels with this kernel mixed into 1) and cube sizes. As this is a continuation of our previous work \cite{informatica} a smaller dataset was targeted due to many hyperspectral unmixing algorithm resources and computation requirements for large datasets.

\par
With the selected sliding windows of size 3x3, 9 pixels were linearly mixed into 1 pixel, including the classes of those pixels. Created dataset cubes were of sizes:
\begin{itemize}
    \item Mixed cube 1 shape: 341 pixels wide, 1059 pixels long with 224 spectral bands of depth.
    \item Mixed cube 2 shape: 341 pixels wide, 1015 pixels long with 224 spectral bands of depth.
    \item Mixed cube 3 shape: 341 pixels wide, 938 pixels long with 224 spectral bands of depth.
\end{itemize}
In previous work \cite{10.1007/978-3-031-63543-4_13}, six classes were used to classify the hyperspectral cubes. With the same sliding window used over the class array, an abundance matrix was generated of the same size as the hyperspectral data but with a third dimension of size 6.

\begin{figure}[t]
\includegraphics[width=\textwidth]{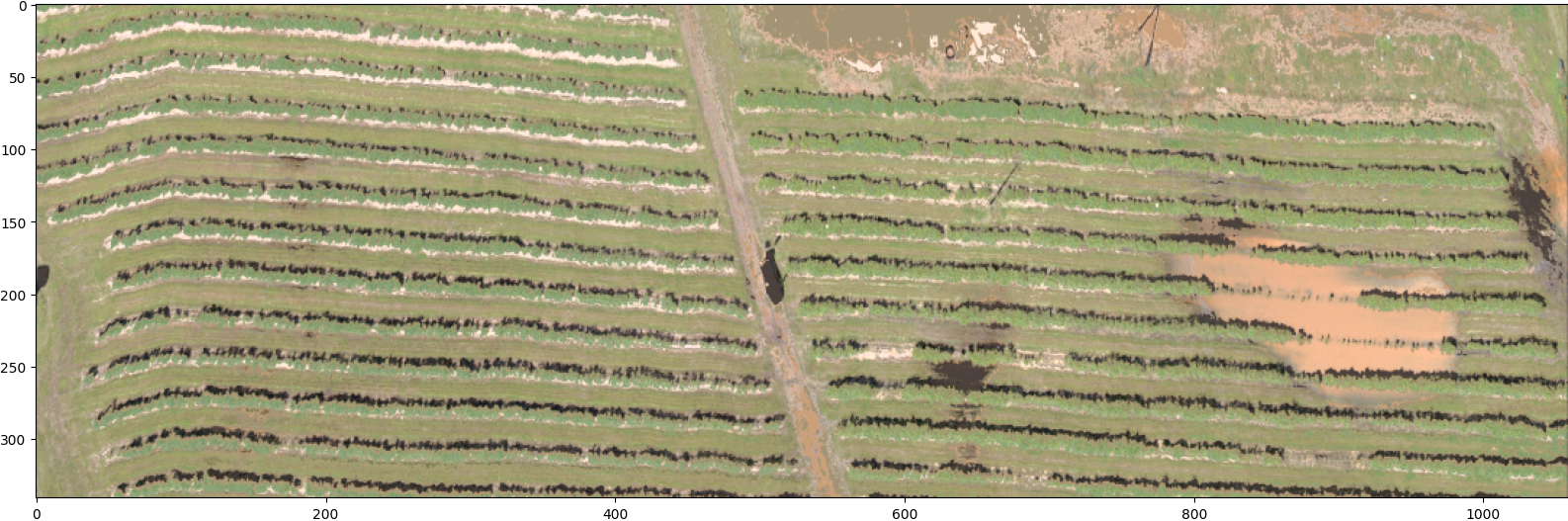}
\caption{Mixed hyperspectral cube RGB representation.} \label{fig:c1_mixed}
\end{figure}

An example RGB representation of data cube 1 is given in fig \ref{fig:c1_mixed}. The RGB representation is computed with data scaled over the whole hyperspectral cube, differences in minimum and maximum value between images distort the final colors, the generated RGB image is a false color image and used only as a convenient visualisation of hyperspectral data cube.

\section{Hyperspectral unmixing using U-Net based architecture}

This section describes our proposed hyperspectral unmixing algorithm based on U-Net model architecture and the datasets used in experimentation and validation. With the popularity of deep learning neural networks and based on existing works in hyperspectral unmixing in agricultural data, an unsupervised deep learning model is a viable solution already used in hyperspectral unmixing.

\subsection{Metrics}
Metrics and losses are used in model training and result comparisons.

\begin{equation}
\label{eq:rmse}
RMSE = \sqrt{\frac{1}{N}\sum_{i=1}^{N}(x_i - \hat x_i)^2},
\end{equation}

\begin{equation}
\label{eq:RE}
L_{RE}(I,\widehat{I})=\frac{1}{W\cdot H}\sum_{i=1}^{H}\sum_{j=1}^{W}(\widehat{I}_{ij}-I_{ij})^2
\end{equation}

\begin{equation}
\label{eq:SAD}
L_{SAD}(I,\widehat{I})=\frac{1}{R}\sum_{i=1}^{R}\arccos\left(\frac{\left \langle I_{i},\widehat{I}_{i} \right \rangle}{ \| I_{i}  \|_{2} \| \widehat{I}_{i}  \|_{2}}\right)
\end{equation}

\begin{equation}
\label{eq:cos}
\cos (\theta ) =   \frac {A \cdot B} {\left\| A\right\| \left\| B\right\| } 
\end{equation}

\begin{itemize}
    \item RMSE (eq: \ref{eq:rmse}) - measures the average difference between values predicted by a model and the actual values. Where $N$ is the number of points that are checked, $i$ - current point index, $x_i$ - actual value, $\hat x_i$ - predicted value;
    \item RE (eq: \ref{eq:RE}) - measures the average difference between model-generated data cube and actual data. Where $W$ is the image width, $H$ image height, $\widehat{I}_{ij}$ - predicted spectral data in pixel with index $i$ $j$, $I_{ij}$ - actual spectral data in pixel with index $i$ $j$.
    \item SAD (eq: \ref{eq:SAD}) - measures the angles between two vectors in multidimensional space. Where $R$ is the number of pixels,  $I_{i}$ - actual data,  $\widehat{I}_{i}$ - predicted data.
    \item Cosine similarity (eq: \ref{eq:cos}) - calculates the dot product of the vectors divided by the product of their lengths. Where $A$ and $B$ are the two input vectors (spectra in this case) to be measured.
\end{itemize}

\subsection{Proposed model architecture}
\begin{figure}[t]
\includegraphics[width=0.8\textwidth]{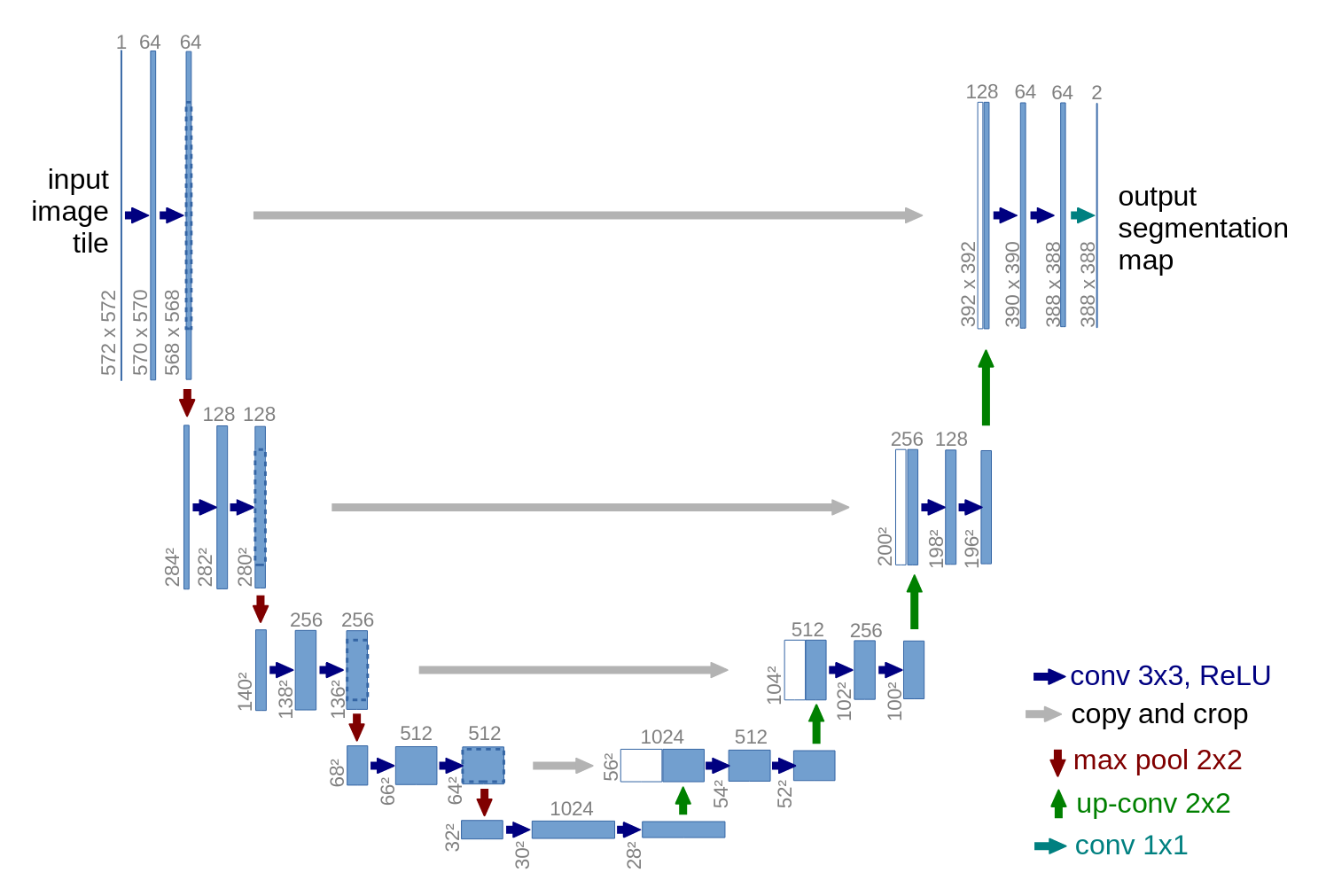}
\caption{U-Net model architecture and layers. Source: \cite{DBLP:journals/corr/RonnebergerFB15}} \label{fig:unet}
\end{figure}

The original U-Net model created by Ronneberger et al. \cite{DBLP:journals/corr/RonnebergerFB15} was used for biomedical image segmentation. Based on our previous work, one of the more common deep learning methods used in hyperspectral unmixing is autoencoder networks. Due to this fact, a base autoencoder architecture from U-Net model was used. The autoencoder compresses data into small latent space to extract features from it at various scales during the compression. Figure \ref{fig:unet} shows the original U-Net model architecture diagram.

\par
To adapt the U-Net model architecture for the hyperspectral unmixing task, a set of changes were made:
\begin{itemize}
    \item Splitting the hyperspectral image into smaller same-size images to reduce the overall size of the model, enabling usage of augmentations (e.g., mirroring and rotations) on the input data and training the model by selecting these image patches in random order.
    \item Addition of cosine similarity loss \ref{eq:cos} was used to encourage the model to extract less similar endmembers.
    \item Splitting the compressed data into endmember and abundance extraction sub-networks.
    \item Fully unsupervised unmixing model is trained on hyperspectral image reconstruction loss.
    \item Ability to provide reference endmembers for more accurate unmixing or faster convergence speeds.
\end{itemize}

\begin{figure}[t]
\centering
\includegraphics[width=0.8\textwidth]{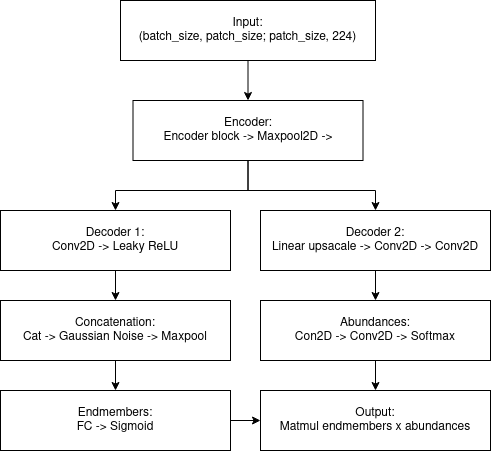}
\caption{The model architecture outline.} \label{fig:arch}
\end{figure}

Figure \ref{fig:arch} shows the model architecture. With the selected batch size and patch size, the model is constructed based on the feature extraction encoder layers, which are split into two parts. First parts extracts the endmembers by compressing the spatial data, second part extracts abundances from the spectral data. To keep the model unsupervised, the data cube is reconstructed in the last layer. To learn the abundances and endmembers, the reconstruction is done using matrix multiplication and not decoder layers directly. Input and output data is the same shape and the model is trained on the reconstruction accuracy.

Full algorithm and experimentation code is open and provided in repository \url{https://github.com/VytautasPau/UAVHyperspectral}.

\subsection{Datasets used for experimentation}
In this section, we analyze freely available hyperspectral datasets that were selected to be used in algorithm performance experimentation as well as including our newly created hyperspectral unmixing dataset:
\begin{itemize}
    \item \textit{DC Mall} \cite{datasets} \cite{dcmall}. An area a part of Washington DC with a size of 1208 x 307 pixels and 191 spectral bands. Created ground truth for classification has these classes: Roofs, Streets, Paths, Grass, Trees, Water, and Shadows.
    \item \textit{Samson} \cite{datasets} Hyperspectral data cube cut to the size of 95 * 95 pixels with 156 spectral bands and 3 different classes.
    \item \textit{Apex} \cite{datasets} Hyperspectral data of size 110 * 110 pixels with 285 spectral bands and 4 different classes.
    \item Cube 1 shape: 1024 pixels wide, 3177 pixels long with 224 spectral bands of depth.
    \item Cube 2 shape: 1024 pixels wide, 3047 pixels long with 224 spectral bands of depth.
    \item Cube 3 shape: 1024 pixels wide, 2815 pixels long with 224 spectral bands of depth.
\end{itemize}

The three open datasets selected were from the original tranformer based hyperspectral unmixing model create by Ghosh et al. \cite{9848995}. Their code included tuned hyper parameters for these datasets as well as a evidance that the model works on these datasets.

\subsection{Evaluation of proposed method}
Model performance was measured using this methodology:
\begin{itemize}
    \item For each dataset the model was trained until the change in reconstruction loss was almost zero, in turn number of training epoch and training time was different for each dataset. Not provided in the results table \ref{tab:res}, training times were mostly dependent on input data size.
    \item Given experimentation results are the best possible achieved with manual hyperparameter tuning. 
    \item For model training reconstruction error (RE, equation \ref{eq:RE}), spectral angle distance (SAD, equation \ref{eq:SAD}) and an additional cosine similarity loss were used as losses. Cosine similarity loss was added to encourage the model to learn endmember that are less similar to each other. If the dataset is suspected to have similar classes this loss should not be used.
    \item For testing the results, root mean squared error (RMSE, equation \ref{eq:rmse}) and SAD metrics averaged over all of the classes were used. In addition hyperspectral cube reconstruction error RE is also provided. 
\end{itemize}

Experimentation results for each of the datasets is provided in table \ref{tab:res}. Our proposed model results were compared to the transformer-based hyperspectral unmixing model created by Ghosh et al. \cite{9848995}. This model was selected due to the code availability, novelty in the hyperspectral unmixing algorithms by using the transformers that showed a higher accuracy compared to other deep neural network model papers available at the time when research was initiated. 

\begin{table}[]
\begin{tabular}{|l|cccl|cccl|}
\hline
\multirow{2}{*}{} &
  \multicolumn{4}{c|}{Proposed model} &
  \multicolumn{4}{c|}{Transformer model} \\ \cline{2-9} 
 &
  \multicolumn{1}{c|}{mRMSE} &
  \multicolumn{1}{c|}{mSAD} &
  \multicolumn{1}{c|}{RE} &
  Epochs &
  \multicolumn{1}{c|}{mRMSE} &
  \multicolumn{1}{c|}{mSAD} &
  \multicolumn{1}{c|}{RE} &
  Epochs \\ \hline
Apex &
  \multicolumn{1}{c|}{\textbf{0.4705}} &
  \multicolumn{1}{c|}{0.1737} &
  \multicolumn{1}{c|}{0.0990} &
  \textit{1001} &
  \multicolumn{1}{c|}{0.5555} &
  \multicolumn{1}{c|}{0.2025} &
  \multicolumn{1}{c|}{0.1048} &
  \textit{1000} \\ \hline
DC &
  \multicolumn{1}{c|}{0.3971} &
  \multicolumn{1}{c|}{0.3764} &
  \multicolumn{1}{c|}{0.0480} &
  \textit{1001} &
  \multicolumn{1}{c|}{\textbf{0.3918}} &
  \multicolumn{1}{c|}{0.3009} &
  \multicolumn{1}{c|}{0.0232} &
  \textit{1000} \\ \hline
Samson &
  \multicolumn{1}{c|}{\textbf{0.4301}} &
  \multicolumn{1}{c|}{0.1507} &
  \multicolumn{1}{c|}{0.0526} &
  \textit{1001} &
  \multicolumn{1}{c|}{0.6031} &
  \multicolumn{1}{c|}{0.2400} &
  \multicolumn{1}{c|}{0.1675} &
  \textit{1000} \\ \hline
\begin{tabular}[c]{@{}l@{}}Blueberry\\ Cube 1\end{tabular} &
  \multicolumn{1}{c|}{\textbf{0.3112}} &
  \multicolumn{1}{c|}{0.2737} &
  \multicolumn{1}{c|}{0.0752} &
  \textit{3001} &
  \multicolumn{1}{c|}{0.4845} &
  \multicolumn{1}{c|}{0.3951} &
  \multicolumn{1}{c|}{0.3012} &
  \textit{1000} \\ \hline
\begin{tabular}[c]{@{}l@{}}Blueberry \\ Cube 2\end{tabular} &
  \multicolumn{1}{c|}{\textbf{0.3740}} &
  \multicolumn{1}{c|}{0.2591} &
  \multicolumn{1}{c|}{0.1263} &
  \textit{3001} &
  \multicolumn{1}{c|}{0.4511} &
  \multicolumn{1}{c|}{0.4012} &
  \multicolumn{1}{c|}{0.2860} &
  \textit{1000} \\ \hline
\begin{tabular}[c]{@{}l@{}}Blueberry\\ Cube 3\end{tabular} &
  \multicolumn{1}{c|}{\textbf{0.3088}} &
  \multicolumn{1}{c|}{0.2214} &
  \multicolumn{1}{c|}{0.0978} &
  \textit{3001} &
  \multicolumn{1}{c|}{0.4232} &
  \multicolumn{1}{c|}{0.3852} &
  \multicolumn{1}{c|}{0.2645} &
  \textit{1000} \\ \hline
\end{tabular}
\caption{Proposed method and transformer networks comparison results on selected datasets and metrics.}
\label{tab:res}
\end{table}

\newpage
\section{Conclusions}

\begin{itemize}
    \item Mean RMSE values achieved with our proposed model were lower than the comparison model on almost all of the datasets. 
    \item RMSE and RE losses are highly correlated, while SAD loss is more unpredictable in turn and may not be suitable to use as the only metric for unmixing tasks.
    \item The transformer model reached the lowest losses earlier than our proposed model and converged faster to local minima.
    \item The transformer model performance after just one epoch was an order of magnitude better than our proposed model.
\end{itemize}

As explained previously, the given results are only the best achieved with manual parameter tuning and, in turn, may not be the best possible results for each of these datasets tested. Some parameter combinations resulted in the model not learning at all. For future works, an automated hyperparameter search is planned to minimize reconstruction error as much as possible.

\newpage

\bibliographystyle{fundam}
\bibliography{citations}


\end{document}